\documentclass[aps,prl,reprint,superscriptaddress,amsmath,amssymb]{revtex4-2}
\usepackage{graphicx}
\usepackage{bm}
\usepackage{xcolor}
\begin{document}
\title{Universal photonic synthetic temporal-mode converter for heterogeneous quantum interconnects}
\author{Joohyung Song}
\affiliation{Department of Electrical and Computer Engineering, University of Maryland, College Park, Maryland 20742, USA}
\author{James I. Kwon}
\affiliation{Department of Mathematics, University of Maryland, College Park, Maryland 20742, USA}
\author{Edo Waks}
\affiliation{Department of Electrical and Computer Engineering, University of Maryland, College Park, Maryland 20742, USA}
\affiliation{Institute for Research in Electronics and Applied Physics, and Joint Quantum Institute, University of Maryland, College Park, Maryland 20742, USA}
\author{Saikat Guha}
\affiliation{Department of Electrical and Computer Engineering, University of Maryland, College Park, Maryland 20742, USA}
\author{Chaohan Cui}
\email{chaohan@umd.edu}
\affiliation{Department of Electrical and Computer Engineering, University of Maryland, College Park, Maryland 20742, USA}
\date{\today}

\begin{abstract}
We propose a synthetic temporal-mode converter (STC) that combines programmable waveform matching with denser time multiplexing for heterogeneous quantum interconnects. Programmable couplers and optical delays reshape known single-photon temporal waveforms through coherent interference between time bins. Conversion to a common single-bin mode concentrates amplitudes from $N$ bins in $\lceil\log_2N\rceil$ mixing stages, simultaneously improving temporal indistinguishability and freeing channel time. Simulations yield mode overlaps above 99.9\% for exponential and Gaussian inputs. For $N=256$ and $0.5\,\mathrm{dB}$ loss per stage, STC gives a 45.6-fold higher photon-pair coincidence rate than temporal gating at equal mode overlap for two exponential inputs. Relative to long pulses, the shortened temporal matched mode can support an approximately 41-fold higher photon-pair coincidence rate when the available time-multiplexing capacity is fully utilized.
\end{abstract}
\maketitle

\paragraph{Introduction---} The distinct requirements of local quantum processing, long-lived quantum memory, and quantum sensing motivate heterogeneous quantum architectures that combine physical platforms with complementary capabilities~\cite{monroe2014large,awschalom2021development,awschalom2025challenges}. Connecting these platforms into a modular network requires quantum interconnects capable of coherently transferring quantum states or distributing entanglement for teleportation and remote quantum gates~\cite{bennett2001remote, cirac1997quantum,duan2010colloquium}. Standard interconnects relies on the linear-optical entanglement swap protocol where the Bell-state measurement on two photons fuses the two matter-photon entanglement~\cite{simon2003robust}. For photons whose temporal envelope is independent of their logical encoding, temporal-mode mismatch reduces the Hong-Ou-Mandel (HOM) interference visibility and limits the heralded entanglement fidelity~\cite{legero2003time, dhara2023entangling}. Such mismatch is a natural challenge for heterogeneous architectures, as different excited-state lifetimes, emitter-cavity couplings, and driving fields of each type of stationary qubit determine distinct temporal amplitudes and phases of photons it entangles with.
\begin{figure*}[t]
\includegraphics[width=\textwidth]{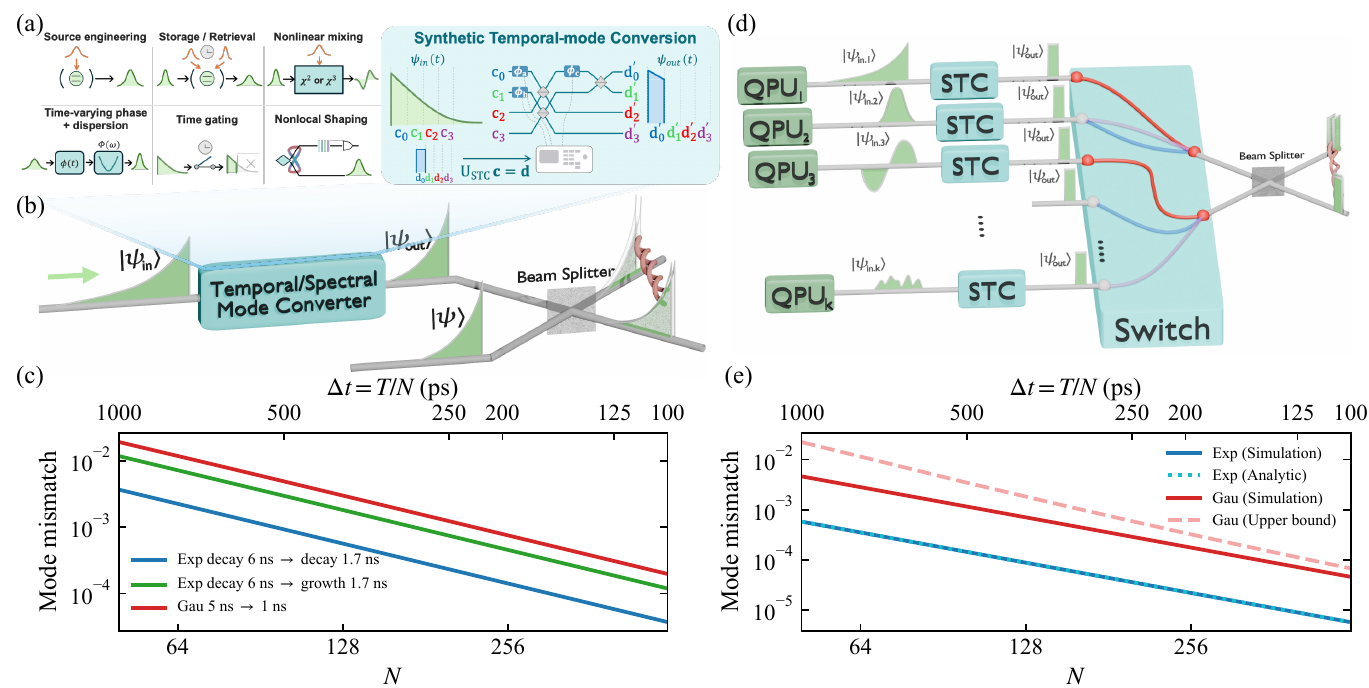}
\caption{\label{fig:figSTC}Temporal-mode conversion for heterogeneous quantum interconnects. (a) Source engineering, quantum pulse shaping, electro-optic temporal-mode shaping, nonlocal pulse shaping, and synthetic temporal-mode conversion. (b) Conversion of one photon to match a reference photon before two-photon interference. (c) Calculated mismatch for exponential decay-to-decay, decay-to-rise, and Gaussian-to-Gaussian conversion. (d) Independently programmed STCs map photons from heterogeneous quantum processing units (QPUs) toward a common single-bin mode for subsequent routing and interference. (e) Mismatch relative to an ideal flat-top single-bin target for exponential and Gaussian inputs, together with the exponential analytic result and the Gaussian bound described in the Supplemental Material~\cite{supplementary}. In (c) and (e), finer binning reduces the conversion error for instantaneous, lossless switching.}
\end{figure*}

Several approaches address temporal-mode mismatch by controlling photon emission or reshaping an incident photon [Fig.~\ref{fig:figSTC}(a)]. Direct source engineering shapes the emitted waveform through time-dependent driving of the emitter--cavity interaction, with the accessible profiles determined by the level structure and coupling parameters~\cite{keller2004continuous,knall2022efficient,10.1038/s41467-017-02552-7}. Conversion after emission can be implemented through controlled storage and retrieval in a quantum-memory interface~\cite{cussenot2025uniting,10.1103/physrevlett.123.133602}, or through nonlinear optical mixing, where the pump envelope and dispersion engineering determine the coupled input and output temporal modes~\cite{10.1088/1367-2630/13/6/065029,allgaier2017highly}. Nonlinear waveform conversion can also compress photon wavepackets coherently~\cite{kielpinski2011quantum}. Electro-optic shaping combines temporal phase modulation with dispersive propagation, using quadratic phase operations for temporal and spectral rescaling~\cite{10.1038/nphoton.2016.228,10.1038/s41566-023-01214-z,krzyzanowski2026quantum}. More general sequences of temporal and spectral phase operations extend this control to arbitrary temporal-mode unitaries~\cite{ashby2020temporal}. Time gating improves photon indistinguishability through temporal postselection, reducing the accepted photon rate \cite{polyakov2011coalescence}. Non-local temporal shaping prepares a heralded single photon with a tailored waveform through temporal modulation and frequency-resolved detection of its time--energy-entangled partner \cite{averchenko2017temporal,bellini2003nonlocal}.

In this Letter, we propose a synthetic temporal-mode converter (STC) for programmable conversion between single-photon waveforms and investigate its temporal-mode shaping capability under hardware imperfections. We extend the time-bin interference architecture of the generalized optical Green Machine~\cite{10.1038/s41467-025-59107-4,10.1103/yvcj-d7pb} to continuous-waveform conversion by explicitly retaining the temporal amplitude and phase profiles within each bin. Discretized input and target modes determine the programmable coupler settings, while the resulting transformation acts on continuous waveform segments coherently aligned by optical delays. This formulation connects the programmed time-bin transformation to the actual output waveform and quantifies the residual mode mismatch arising from unresolved variation within each bin. Using exponential and aussian waveforms, we evaluate how bin width, truncation, insertion loss, and finite switching response jointly determine output mode profile and transmission. We further show numerically that conversion of long photon waveforms toward a common single-time-bin mode [Fig.~\ref{fig:figSTC}(d)] improves temporal mode matching while reducing channel time-slot occupancy, enabling denser time multiplexing for high-fidelity heterogeneous quantum interconnects.

\paragraph{Composing the STC---} We begin with the STC construction between known input and target waveforms. We represent the single-photon state in continuous waveforms by
\begin{equation}
|\psi_\mu\rangle=\int_{-\infty}^{\infty}\psi_\mu(t)\hat a^\dagger(t)\,dt\,|\mathrm{vac}\rangle,
\label{eq:continuousstates}
\end{equation}
with $\int|\psi_\mu(t)|^2dt=1$, and $\hat a^\dagger(t)$ the time-domain photon creation operator. Here, $\mu$ labels the input ($\mathrm{in}$) or target ($\mathrm{tar}$) mode. 

To construct the transformation in a discrete time-bin grid, we select a processing window $W=[t_0,t_0+T)$ and divide it into $N$ bins of width $\Delta t=T/N$. Bin $j$ occupies $W_j=[t_j,t_j+\Delta t)$, where $t_j=t_0+j\Delta t$ and $j=0,\ldots,N-1$. The construction uses only the waveforms within $W$, which retains the capture probability $\eta_W=\int_W|\psi_{\mathrm{in}}(t)|^2dt$.

We then approximate each waveform in orthonormal flat-top bin modes with creation operators $\hat a_j^\dagger=(\Delta t)^{-1/2}\int_{W_j}\hat a^\dagger(t)\,dt$. Each mode has amplitude $1/\sqrt{\Delta t}$ inside its bin and zero outside. The normalized input and target coefficients of each time bin are
\begin{equation}
c_j=\frac{1}{\mathcal N_{\mathrm{in}}}\int_{W_j}\psi_{\mathrm{in}}(t)\,dt,\quad
 d_j=\frac{1}{\mathcal N_{\mathrm{tar}}}\int_{W_j}\psi_{\mathrm{tar}}(t)\,dt,
\label{eq:binamplitudes}
\end{equation}
where $\mathcal N_{\mu}=\bigl[\sum_j|\int_{W_j}\psi_{\mu}(t)\,dt|^2\bigr]^{1/2}$ normalizes each coefficient vector. These coefficients define discretized bin states $|\psi_{\bm c}\rangle=\sum_jc_j\hat a_j^\dagger|\mathrm{vac}\rangle$ and $|\psi_{\bm d}\rangle=\sum_jd_j\hat a_j^\dagger|\mathrm{vac}\rangle$ that approximate the continuous input and target within $W$~\cite{brecht2015photon}. With coefficient vectors $\bm c$ and $\bm d$, the converter requires an $N\times N$ unitary satisfying $U_{\mathrm{STC}}\bm c=\bm d$, or equivalently
\begin{equation}
\hat U_{\mathrm{STC}}|\psi_{\bm c}\rangle=|\psi_{\bm d}\rangle.
\label{eq:coarsemap}
\end{equation}

A full singular value decomposition (SVD) of the outer product $\bm d\bm c^\dagger=A\Sigma B^\dagger$ gives $U_{\mathrm{STC}}=AB^\dagger$, which can be realized by programming couplers and optical delays in the generalized temporal Green Machine architecture~\cite{10.1038/s41467-025-59107-4,10.1103/yvcj-d7pb}. Each known input--target pair determines a converter setting.

Although $U_{\mathrm{STC}}$ is constructed from flat-top bin states, the physical converter mixes continuous waveform segments that retain their variation within each bin. We write the retained input as $x_j(s)=\psi_{\mathrm{in}}(t_j+s)/\sqrt{\eta_W}$, where $s$ measures time from the start of bin $j$ and $0\leq s<\Delta t$. In the ideal, lossless model, the control settings are constant within each bin and change instantaneously at its boundaries. Optical delays align corresponding times $s$ across bins. The output segment is then $y_j(s)=\sum_kU_{jk}x_k(s)$, where $U_{jk}$ is the matrix element connecting input bin $k$ to output bin $j$. This matrix maps the time-bin-approximated input state to targeted output states exactly, but the continuous waveforms of the output state can still differ from the continuous target state, leaving residual temporal-mode mismatch.

To resolve this mismatch numerically, we divide each bin into $M$ numerical cells of width $h=\Delta t/M$, giving $K=NM$ cells across $W$. Sampling the input and target wavefunctions on this common grid and normalizing each coefficient vector gives $\bm\psi_{in}^{(K)}$ and $\bm\psi_{\mathrm{tar}}^{(K)}$. Ordering these entries bin by bin, the transformation is $\widetilde U^{(K)}=U_{\mathrm{STC}}\otimes I_M$, where $I_M$ is the identity matrix to broadcast the matrix elements $U_{jk}\delta_{m\ell}$ to mix only corresponding cells $m,\ell$ across $j$ and $k$ bins. We check overlap convergence by increasing $M$ at fixed $N$ and $\Delta t$, leaving the physical converter unchanged.

\begin{figure}[t]
\includegraphics[width=\columnwidth]{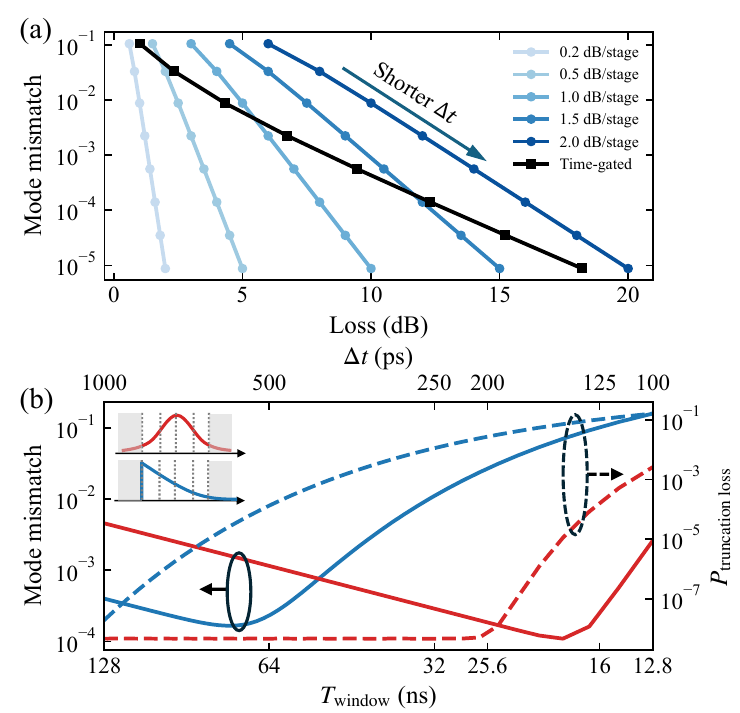}
\caption{\label{fig:figSTB+Loss} Trade-off between additional loss and mode mismatch in STC. (a) Mutual output mismatch for two exponential inputs, $\tau=1.7$ and $6\,\mathrm{ns}$, after independent designed STC into a single time-bin versus direct temporal gating. STC curves use $3\leq n\leq10$ and the indicated per-stage loss. The horizontal axis shows the attenuation in decibels associated with the geometric-mean single-photon transmission $\sqrt{\eta_1\eta_2}$. Here $\eta_1$ and $\eta_2$ are the transmission of the two STCs. (b) Finite-window conversion for $N=128$ and input intensity FWHM $5\,\mathrm{ns}$. Solid curves give conditional mode mismatch after renormalization of the truncated waveforms. Dashed curves give the truncation probability $1-\eta_W$. Blue and red correspond to exponential and Gaussian inputs, respectively. The upper and lower axes are related by $T=N\Delta t$.}
\end{figure}

\paragraph{Performance benchmarking---} We assess mode matching after STC and its tradeoff with additional photon loss. For ideal switching, phase, and tunable beamsplitter setting without loss, the normalized states within $W$ give the target mismatch
\begin{equation}
\epsilon_{\mathrm{tar}}=1-|\bm\psi_{\mathrm{tar}}^{(K)\dagger}\widetilde U^{(K)}\bm\psi_{in}^{(K)}|^2.
\label{eq:modemismatch}
\end{equation}

For two-photon interference, let $|\psi_{\mathrm{out},1}\rangle$ and $|\psi_{\mathrm{out},2}\rangle$ denote the two normalized output states meeting at the beamsplitter for HOM interference. Assuming pure single photons with matched spatial, polarization, and carrier-frequency degrees of freedom, the temporal-mode-determined mode-matching visibility is \mbox{$V=|\langle\psi_{\mathrm{out},1}|\psi_{\mathrm{out},2}\rangle|^2$}, and the mutual mode mismatch is $1-V$. Throughout this Letter, we evaluate this overlap on the $K=NM$ grid. Insertion losses and truncation-induced losses are accounted separately in transmission probabilities.

We uses Gaussian and exponentially decaying waveforms as representative inputs,
\begin{equation}
\psi_{\mathrm G}(t;\sigma)=\frac{e^{-t^2/(4\sigma^2)}}{(2\pi)^{1/4}\sqrt{\sigma}},~
\psi_{\mathrm E}(t;\tau)=\frac{e^{-t/(2\tau)}}{\sqrt{\tau}}\Theta(t),
\label{eq:profiles}
\end{equation}
with $\sigma$ the Gaussian intensity standard deviation, $\tau$ the exponential intensity decay time, and $\Theta$ the Heaviside function. Figure~\ref{fig:figSTC}(c) compares three conversion circumstances. In the first scenario, we shorten an exponential decay from $\tau=6\,\mathrm{ns}$ to $1.7\,\mathrm{ns}$, using timescales motivated by trapped-ion and diamond silicon-vacancy emission~\cite{PhysRevA.77.060501,rogers2014multiple}. In the second scenario, we convert a decaying exponential ($\tau=6\,\mathrm{ns}$) to a rising exponential ($\tau=1.7\,\mathrm{ns}$), relevant to time-reversed absorption protocols~\cite{aljunid2013excitation,leong2016time}. The third scenario simulates compressing a Gaussian pulse from an intensity FWHM of $5\,\mathrm{ns}$ ($\sigma=2.12$) to $1\,\mathrm{ns}$ ($\sigma=0.42$). In all three cases, the conversion mismatch scales approximately as $N^{-2}$, or equivalently as $(\Delta t)^2$, for a fixed processing window $T$.

Then we investigate performances of STC in heterogeneous quantum interconnects where independently programmed STCs can map each known input waveform toward a common single-time-bin mode for routing and interference [Fig.~\ref{fig:figSTC}(d)], providing a universal temporal-mode interface supporting the universal quantum bus~\cite{englund2026bus} and enabling denser time multiplexing. We choose a flat-top mode in the first output bin $j=0$, with amplitude $1/\sqrt{\Delta t}$ inside that bin and zero outside as the target. Using the input coefficients $\bm c$ defined above, we program $\hat U_{\mathrm{STC}}|\psi_{\bm c}\rangle=\hat a_0^\dagger|\mathrm{vac}\rangle$. For a known input spanning $N$ bins, recursive pairwise couplings concentrate its bin amplitudes into one bin in $n=\lceil\log_2N\rceil$ mixing stages after padding to a power of two.

The transformation leaves mismatch from the continuous temporal shape within each bin. Applying the Cauchy-Schwarz inequality shows the minimum mismatch is $\epsilon_{\mathrm{flat}}^{\min}=1-(\Delta t)^{-1}\sum_j|\int_0^{\Delta t}x_j(s)\,ds|^2$. Bounding each segment's deviation from its bin average gives
\begin{equation}
\epsilon_{\mathrm{flat}}^{\min}\leq
\frac{\Delta t^2}{\pi^2}\sum_j\int_0^{\Delta t}|x'_j(s)|^2ds.
\label{eq:flatbound}
\end{equation}
Here $x'_j(s)=dx_j(s)/ds$~\cite{supplementary}. For smooth envelopes, the bound decreases quadratically with the bin width $\Delta t$. The input segments $x_j(s)$ are normalized after truncation to $W$, so $\epsilon_{\mathrm{flat}}^{\min}$ measures the mismatch of the retained waveform with the additional loss led by the truncation $1-\eta_W$. Therefore, this mismatch accounts the converted output across the full window $W$, including any residual amplitude outside the target bin.

An extra switch can route only the designated target bin, $j=0$ here, into the shared channel and reject the other output bins. This leaves the remaining time slots available for converted photons from other sources, enabling denser time multiplexing. Furthermore, rejecting these bins removes their contribution to the transmitted photon's mode mismatch, thus ensuring a slightly reduced mode mismatch while sacrificing slight extra loss. For an ideal switch, the transmitted fraction of the retained photon probability is $\eta_{\mathrm{bin}}=\int_0^{\Delta t}|y_0(s)|^2ds$. We calculate the transmitted photon's mode overlap after renormalizing its waveform to $y_0(s)/\sqrt{\eta_{\mathrm{bin}}}$.

For exponential inputs, the bin segments share the same profile and differ only in amplitude. The programmed unitary therefore transfers all retained photon probability into the first output bin. The target mode mismatch for this exponential input is
\begin{equation}
\epsilon_{\mathrm{flat,E}}=1-\frac{4\tau}{\Delta t}\tanh\!\left(\frac{\Delta t}{4\tau}\right)
\simeq\frac{\Delta t^2}{48\tau^2}.
\label{eq:expfidelity}
\end{equation}
The expansion applies for $\Delta t\ll\tau$. For a Gaussian input, we evaluate the minimum target mismatch directly from the bin integrals above Eq.~(\ref{eq:flatbound})~\cite{supplementary}. Figure~\ref{fig:figSTC}(e) shows that the target mismatch decreases with finer binning for both profiles, falling below $10^{-3}$ at the finest resolution shown. The converted exponential and Gaussian outputs also have a mutual mode-matching visibility above 99.9\%~\cite{supplementary}.

\begin{figure}[t]
\includegraphics[width=\columnwidth]{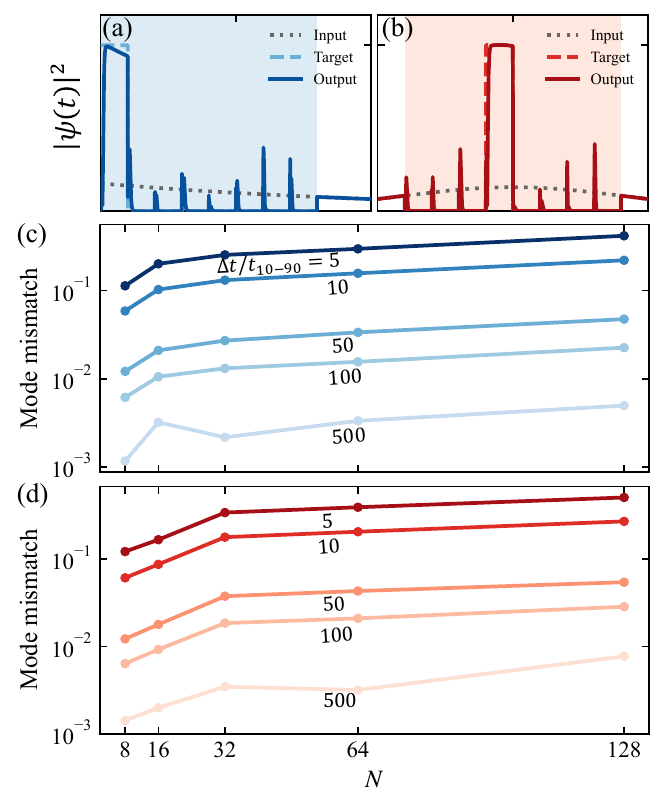}
\caption{\label{fig:figNonideal} STC with finite switching response. (a),(b) Representative output intensities (solid) and target intensities (dashed) for exponential and Gaussian inputs, with shading indicating the processing window. (c),(d) Reported normalized output mismatch for the respective inputs as a function of $N$, at $\Delta t=0.5\,\mathrm{ns}$. The curves are labeled by $\Delta t/t_{10\text{--}90}$, with ratios from 5 to 500 corresponding to increasingly rapid switching.}
\end{figure}

To quantify the associated photon loss, we assume that each traversed STC stage attenuates all time bins equally by $L_{\mathrm{stage}}$ decibels (dB). The normalized waveform is therefore preserved, while the transmission probability is multiplied by $\eta_{\mathrm{ins}}=10^{-nL_{\mathrm{stage}}/10}$. The losses from input-window truncation and rejecting residual bins enters the overall STC loss budget along with  the device insertion loss. As a benchmark, temporal gating transmits only the part of an input pulse within a chosen time interval and rejects the rest. For an exponential pulse starting at $t=0$, a gate covering $0\leq t<w$ transmits a fraction $1-e^{-w/\tau_i}$ of photon $i$. Setting $w=\Delta t$ gives the same normalized waveform as ideal STC conversion into one bin. The STC instead transmits $\eta_{\mathrm{ins},i}(1-e^{-T/\tau_i})$ by combining amplitude from the full processing window $T$.

Figure~\ref{fig:figSTB+Loss}(a) compares mutual output mode mismatch $1-V$ against photon loss for STC and temporal gating, using exponential inputs with $\tau_1=1.7\,\mathrm{ns}$ and $\tau_2=6\,\mathrm{ns}$. The STC curves sweep the stage count from $n=3$ to 10, changing $N=2^n$ and $\Delta t=T/N$ at fixed $T$, with a separate curve for each indicated loss per stage. The gating curve varies the gate width $w=\Delta t$. The horizontal axis reports the attenuation in dB associated with $\sqrt{\eta_1\eta_2}$, where $\eta_i$ is the transmission probability for photon $i$. For $T=50\,\mathrm{ns}$, $N=256$, $\tau_1=1.7\,\mathrm{ns}$, and $\tau_2=6\,\mathrm{ns}$, STC and temporal gating with $w=T/N$ both give $V=0.999859$. At this equal mode overlap, $0.5\,\mathrm{dB}$ loss per STC stage gives a pair transmission probability of 0.15845, compared with 0.003476 for gating, a 45.6-fold advantage~\cite{supplementary}. This advantage decreases as the stage loss increases. At $2\,\mathrm{dB}$ per stage, the accumulated loss makes STC less efficient than gating [Fig.~\ref{fig:figSTB+Loss}(a)].

The processing window introduces a second tradeoff. Figure~\ref{fig:figSTB+Loss}(b) varies $T$ at fixed $N=128$, corresponding to seven concentration stages. A longer window retains more photon probability but also widens the bins, since $\Delta t=T/N$, and therefore limits the mode-matching visibility. A shorter window improves temporal resolution while rejecting more of the input. The dashed curves show that an exponential tail requires a longer window than a centered Gaussian to retain the same probability. The solid curves give the normalized mode mismatch that accompanies this truncation loss.

Finite switching speed adds another source of conversion error. To model the limited bandwidth of the control electronics and electro-optic modulators (EOMs), we apply a single-pole RC filter to the control phase of each $2\times2$ mixing operation in all $n=\lceil\log_2N\rceil$ STC stages, using the sine--cosine fractal decomposition~\cite{basani2023self,10.1103/yvcj-d7pb}. A measured device response can replace this simplified illustrative model. For an RC time constant $\tau_{\mathrm{RC}}$, the normalized step response is $r(t)=1-e^{-t/\tau_{\mathrm{RC}}}$ for $t\geq0$, with a 10--90\% rise time $t_{10\text{--}90}\simeq2.197\tau_{\mathrm{RC}}$.

Figure~\ref{fig:figNonideal} shows how finite switching speed affects the output waveforms and mode mismatch at a bin width of $\Delta t=0.5\,\mathrm{ns}$. The ratios $\Delta t/t_{10\text{--}90}=5$ to 500 correspond to rise times from 100 to $1\,\mathrm{ps}$. A finite response smooths the switching boundaries and sends amplitude into unwanted bins. The resulting mismatch depends on the circuit size and control sequence as well as the switching speed. At a ratio of 500, the mismatch remains below 1\% for both profiles at every plotted $N$. At a ratio of 100, it exceeds 1\% for some values of $N$. Timing fluctuations impose a further constraint, considered separately in the Supplemental Material~\cite{supplementary}.

Converting photons into single-bin outputs also reduces the channel time occupied by each photon. A reference interval $T=N\Delta t$, previously allocated to one long wavepacket, can then accommodate up to $N$ converted photons from independent sources. Each source uses a local STC, and the designated outputs are routed into successive time bins of the shared optical channel. The usable slot count depends on guard intervals, routing, and sufficient channel bandwidth to transmit the compressed pulses. Sharing the optical link can also allow reuse of its optical phase-stabilization system in protocols that require a common phase reference, including coherent-state-mediated memory entanglement~\cite{azuma2012coherent,cui2025coherent}.

To quantify the benefit in multiplexed channel rate, we compare the photon-pair coincidence rates with and without STC, counting pairs for which both photons arrive at the central node. The reference rate $R_{\mathrm{long}}$ uses unconverted, originally mode-matched long wavepackets, with one photon pair launched per interval $T$. Both schemes use the same optical links and must meet the same required mode-overlap threshold. Let $\eta_{\mathrm{STC},i}$ be the transmission probability into the designated output bin in arm $i$. If parallel sources and sufficient converter and receiver throughput keep all $N$ output slots occupied, the higher time-multiplexing capacity allows $N$ photon pairs to be launched per interval $T$. Writing the joint-arrival rate with STC as $R_{\mathrm{STC}}$, the photon-pair coincidence rate gain is
\begin{equation}
\frac{R_{\mathrm{STC}}}{R_{\mathrm{long}}}
=N\eta_{\mathrm{STC},1}\eta_{\mathrm{STC},2},
\label{eq:rategain}
\end{equation}
For $N=256$, eight STC stages with $0.5\,\mathrm{dB}$ loss per stage give $\eta_{\mathrm{STC},1}=\eta_{\mathrm{STC},2}\simeq0.398$, yielding an approximately 41-fold photon-pair coincidence rate gain. The increased rate enabled by the higher time-multiplexing capacity thus outweighs the insertion loss under these conditions. The two benchmarks show gains in photon-pair coincidence rate when the outputs meet the same required mode-overlap threshold. For the exponential inputs in Fig.~\ref{fig:figSTB+Loss}(a), STC gives a 45.6-fold higher pair coincidence rate than gating at equal mode overlap and equal input pair launch rates. More general scheduling and loss conditions are given in the Supplemental Material~\cite{supplementary}.

\paragraph{Discussion---} The proposed STC combines temporal-mode matching functionality with more efficient use of a shared optical channel by coherently reshaping known single-photon waveforms. The same linear-optical mode transformation could also reshape more complex photonic resource states occupying a known transform-limited temporal mode, such as pulsed squeezed states~\cite{cui2021squeezing}. Realizing these benefits requires low insertion loss at the switching speeds needed to address individual time bins. Improved EOM coupling, ultra-low-loss delay lines, and removal of unnecessary optical traversals offer routes toward sub-decibel stage loss, as discussed for the optical Green Machine and its generalized architecture~\cite{10.1038/s41467-025-59107-4,10.1103/yvcj-d7pb}. Achieving this loss budget at the required switching bandwidth remains an implementation challenge.

Reversing the conversion stretches short photons into longer temporal modes. This could connect photons from quantum sensors, quantum dots, or entangled-photon sources to narrowband quantum memories that interact with quasi-continuous-wave laser pulses. Nanophotonic diamond silicon-vacancy memories interfaced with laser-derived time-bin qubits provide an experimental example of such a memory interface~\cite{knaut2024entanglement}. One prospective application is loading memories from zero-added-loss multiplexed (ZALM) entangled-photon sources~\cite{chen2023zalm,shapiro2024zalm}, with temporal stretching complementing carrier-frequency conversion.

\emph{Acknowledgments.---}
This material is based upon work supported by the Defense Advanced Research Projects Agency (DARPA) under Agreement No. HR0011-26-9-E114. Approved for public release; distribution is unlimited. J.S., E.W, S.G., and C.C also acknowledge support from the NSF Engineering Research Center for Quantum Networks (CQN, EEC-1941583).

\emph{Data Availability.---}The data and codes that support the findings of this Letter are available upon request.

\bibliography{STCref_arXiv}
\end{document}